\documentclass[aps,prd,showpacs,floatfix,superscriptaddress,nofootinbib,10pt,twocolumn]{revtex4}
\usepackage{graphicx, epsfig}
\usepackage[dvipsnames]{xcolor}
\usepackage{hyperref}
\usepackage{mathrsfs}
\usepackage{amsmath}
\usepackage{enumerate}
\usepackage{bm}

\usepackage{float}
\usepackage{ulem}

\begin{document}

\title{Gravitational Wave From Axion-like Particle Inflation}

\author{Wei Cheng}
\affiliation{School of Science, Chongqing University of Posts and Telecommunications, Chongqing 400065, P. R. China}
\affiliation{State Key Laboratory of Theoretical Physics, Institute of Theoretical Physics, Chinese Academy of Sciences, Beijing, 100190, P.R. China}
\author{Tao Qian}
\affiliation{Department of Physics, Nanjing University, 22 Hankou Road, Nanjing 210093, P. R. China}
\author{Qing Yu}
\affiliation{Department of Physics, Chongqing University, Chongqing 401331, People's Republic of China}
\author{Hua Zhou}
\affiliation{Department of Physics, Chongqing University, Chongqing 401331, People's Republic of China}
\author{Rui-Yu Zhou\footnote{Corresponding author}}
\email{zhoury@cqupt.edu.cn}
\affiliation{School of Science, Chongqing University of Posts and Telecommunications, Chongqing 400065, P. R. China}
\affiliation{Department of Physics, Chongqing University, Chongqing 401331, People's Republic of China}

\date{\today}

\begin{abstract}{
In this paper, we investigate the Axion-like Particle inflation by applying the multi-nature inflation model, where the end of inflation is achieved through the phase transition (PT). The events of PT should not be less than $200$, which results in the free parameter $n\geq404$.
Under the latest CMB restrictions, we found that the inflation energy is fixed at $10^{15} \rm{GeV}$.
Then, we deeply discussed the corresponding stochastic background of the primordial gravitational wave (GW) during inflation.
We study the two kinds of $n$ cases, i.e., $n=404, 2000$.
We observe that the magnitude of $n$ is negligible for the physical observations, such as $n_s$, $r$, $\Lambda$, and $\Omega_{\rm{GW}}h^2$.
In the low-frequency regions, the GW is dominated by the quantum fluctuations,
and this GW can be detected by Decigo at $10^{-1}~\rm{Hz}$.
However, GW generated by PT dominates the high-frequency regions, which is expected to be detected by future 3DSR detector.
}
\end{abstract}

\maketitle

\section{Introduction}

With the first discovery of gravitational wave (GW) from a binary neutron star (BNS) system by the Laser Interferometer Gravitational Wave Observatory
(LIGO)~\cite{TheLIGOScientific:2014jea} and by the Virgo detector~\cite{TheVirgo:2014hva},
the primary GW from inflation is increasingly the focus of theorists and experimentalists.
Inflationary scenario not only explain the problem in the hot big bang universe gracefully, i.e., the
flatness puzzles, the horizon puzzles, and the monopole puzzles, (refer to~\cite{Guth:1980zm,Albrecht:1982wi,Linde:1981mu,revlinde,liddle,riotto,mukhanov,nojirio,nojirioo}),
but also predict the nearly scale-invariance which is strongly supported by the anisotropies of cosmic microwave background (CMB) radiation~\cite{Akrami:2018odb}.

Typically, in the inflationary models with broken slow-roll conditions~\cite{Lerner:2009xg,Zheng:2014dra,Munoz:2014eqa,Aravind:2015xst,Cheng:2018ajh,Cheng:2018axr},
the primary GW only comes from the quantum fluctuations (QF) during inflation~\cite{Matarrese:1997ay,Bartolo:2007vp,Biagetti:2013kwa,Biagetti:2014asa},
and its intensity is related to the tensor-to-scalar ratio $r$ of inflation~\cite{Liu:2015psa,Boyle:2005se}.
However, the $r$ predicted by the theory, in this case, is so small that a weak GW will be generated~\cite{Starobinsky:1980te},
which brings great challenges for experimental detection~\cite{Wang:2018caj}.

Alternatively, in the literatures~\cite{Linde:1990gz,Adams:1990ds,Ashoorioon:2015hya}, a new way to end inflation has been proposed, i.e., the inflationary end of the bubble nucleation method.
This method also has the natural function of reheating the universe through the collision of true vacuum bubble walls and convert it to radiation.
Here are two scalar fields in this inflationary models, one of the fields is responsible for driving the inflation of the universe,
whereas the other field can indeed percolate the true vacuum from the meta-stable one, and complete the phase transition (PT) and end inflation simultaneously.
In this case, the source of GW includes quantum fluctuations during inflation as well as the PT that will produce considerable GW~\cite{Caprini:2015zlo,Binetruy:2012ze,Ashoorioon:2008nh}.
This makes it possible to detect GW during inflation.

In this work, a singlet Axion-like Particle (ALP) is applied to drive cosmic inflation and to terminate the inflation through the PT simultaneously,
and we then investigate in detail the GW produced by the ALP inflation.
The multi-natural inflation (MNI)~\cite{Czerny:2014wza,Takahashi:2019qmh} based on natural inflation (NI)~\cite{Adams:1992bn} can increase the feasible parameter space of the NI that is shrink by the latest CMB data~\cite{Akrami:2018odb}.
The inflation of MNI has a good property, shift symmetry, allows an ALP to act as the inflaton~\cite{Daido:2017wwb,Takahashi:2019qmh}.
Besides, it also keeps the flatness of potential energy, which is essential for driving the expansion of the universe and
generating density perturbations. Moreover, MNI can provide multiple false vacuums due to the multiple cosine functions in the potential of MNI.
With the slow-roll inflation proceeds, the inflation first falls into the false vacuum, the vacuum bubbles then collide with each other,
and the inflation finally transits into the true vacuum.
This provides the possibility for MNI of ALP to complete cosmic inflation and to end inflation through PT simultaneously.
We calculate the GW for the period of ALP inflation that will consist of two parts, namely, the QF GW and the PT GW.

The remaining parts of the paper are organized as follows.
In Sec.~\ref{ALPIf}, we present the Multi-NI model and study the inflation exit from the PT.
In Sec.~\ref{GWsIf}, the GW from both QF and PT are in detail discussed.
Sec.~\ref{Summary} is a summary for this work.

\section{Axion-like particle inflation}\label{ALPIf}

\begin{figure*}[ht!]
\begin{center}
\includegraphics[width=0.5\textwidth]{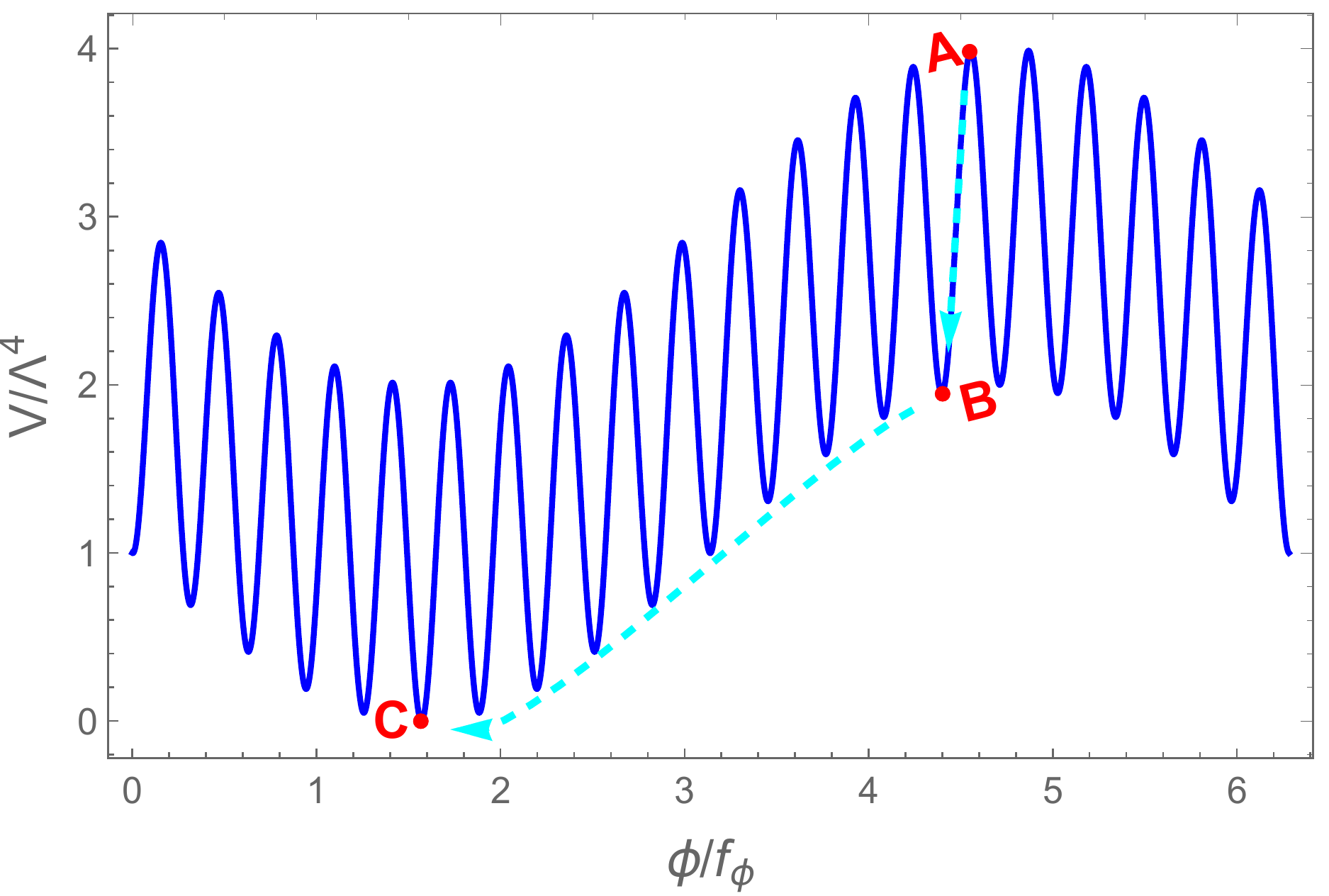}
\end{center}
\caption{Schematic diagram of ALP inflation and PT with $n=20$. }
\label{Fig:Fig_V}
\end{figure*}

The minimal axion-like particle inflation potential derived from a multi-natural inflation, is defined as follows~\cite{Czerny:2014wza,Takahashi:2019qmh}:
\begin{eqnarray}\label{EQ:P_Ainf}
V(\phi)=\Lambda^4[\cos(\frac{\phi}{f_\phi}+\theta)-\frac{\kappa}{n^2} \cos(n \frac{\phi}{f_\phi})]+C
\end{eqnarray}
where $\Lambda$ is related to the inflation scale, $f_\phi$ represents the decay constant.
We set the a relative phase $\theta=\pi/2$ for creating false vacua between the maximum and minimum values of adjacent potential,
and take the coefficient $\kappa=n^2$ for equalizing the energy difference between two adjacent vacua.
For the sake of discussion, we require the minimum of the potential at $\phi=\pi/2$, thus the rational number $n$ is a multiple of $4$.
And $n$ shall determine the tunneling events of the PT which are more than $200$ with a GUT scale inflation~\cite{Ashoorioon:2008nh}. Thus the rational number $n$ should be at least $404$. $C=2\Lambda^4$ is a constant that shifts the minimum of the potential to zero.

For the convenience of presentation, we take a small value for $n=20$ and show the ALP inflation potential in Fig.~\ref{Fig:Fig_V}, in which the dashed cyan with arrows represents the direction of inflation and phase transitions. The inflaton first rolls to the first false vacuum point B from the highest point A of potential then reaches the lower vacuum point through a phase transition, and finally reaches the true vacuum point C.

During the slow-roll inflation, the kinetic energy of inflaton is much less than that of the potential energy, and the change rate of kinetic energy should much less than that of the cosmic expansion, which can be characterized by the so called slow-roll conditions:
\begin{equation}
\varepsilon(\phi) \equiv \frac{{M_{pl}}^2}{2}\left(\frac{V'(\phi)}{V(\phi)}\right)^2 \ll 1,\\
\quad \eta(\phi) \equiv {M_{pl}}^2\left(\frac{V''(\phi)}{V(\phi)}\right)\ll 1,
\end{equation}
where the prime denotes the derivative with respect to $\phi$.

Generally, the inflation will be finished with any one of the conditions breaks, i.e. $\varepsilon=1$ or $\mid\eta\mid=1$. However, we need the inflation exited with the bubble nucleation, which means that the above conditions are still held at the beginning of PT, i.e., $\varepsilon(\phi_B)\ll 1$ and $\eta(\phi_B)\ll 1$ with $\phi_B$ being the field value of the first vacuum, which will bring a lower limit for the decay constant $f_\phi$ with a fix $n$. As an example, we will show two cases as follows:
\begin{align}
f\gg& \sqrt{4.39742\times10^{31}},~~~\rm{for}~~n=404;\\
f\gg& \sqrt{7.10613\times10^{30}},~~~\rm{for}~~n=2000.
\end{align}

There are two observable quantities, i.e., scalar spectral index $n_s = 1-6\varepsilon+2\eta$ and the tensor-to-scalar ratio $r = 16\varepsilon$, which are tightly constrained by the Planck data combined with CMB observations as~\cite{Akrami:2018odb}:
\begin{align}
n_s &= 0.9649 \pm 0.0042, \\
r &< 0.10 ~~~(95\% {\rm CL}).
\end{align}
For calculating them at the moment of horizon-crossing, we need to know the value of the field where inflation starts, $\phi_*$, which are usually determined by applying for the e-folding number. Numerical calculation shows that the contributions of the e-folding number from the highest potential energy point $\phi_A$ to the first vacuum point $\phi_B$ are negligible, thus we're approximating $\phi_*$ around the highest potential energy point $\phi_A$.

There is another important experimental observation, i.e. amplitude of scalar fluctuations $\Delta_\mathcal{R}^2$,
\begin{equation}
\Delta_\mathcal{R}^2 = \frac{1}{24 \pi^2 M_{\rm pl}^4}\frac{V(\phi_A)}{\varepsilon(\phi_A)}.
\end{equation}
where $\Delta_{\mathcal R}^2 \simeq e^{3.098} \times 10^{-10} \simeq 2.2 \times 10^{-9}$~\cite{Ade:2015lrj}, which will give out the inflation scale $\Lambda$.

\begin{figure*}[t]
\begin{center}
\includegraphics[width=0.45\textwidth]{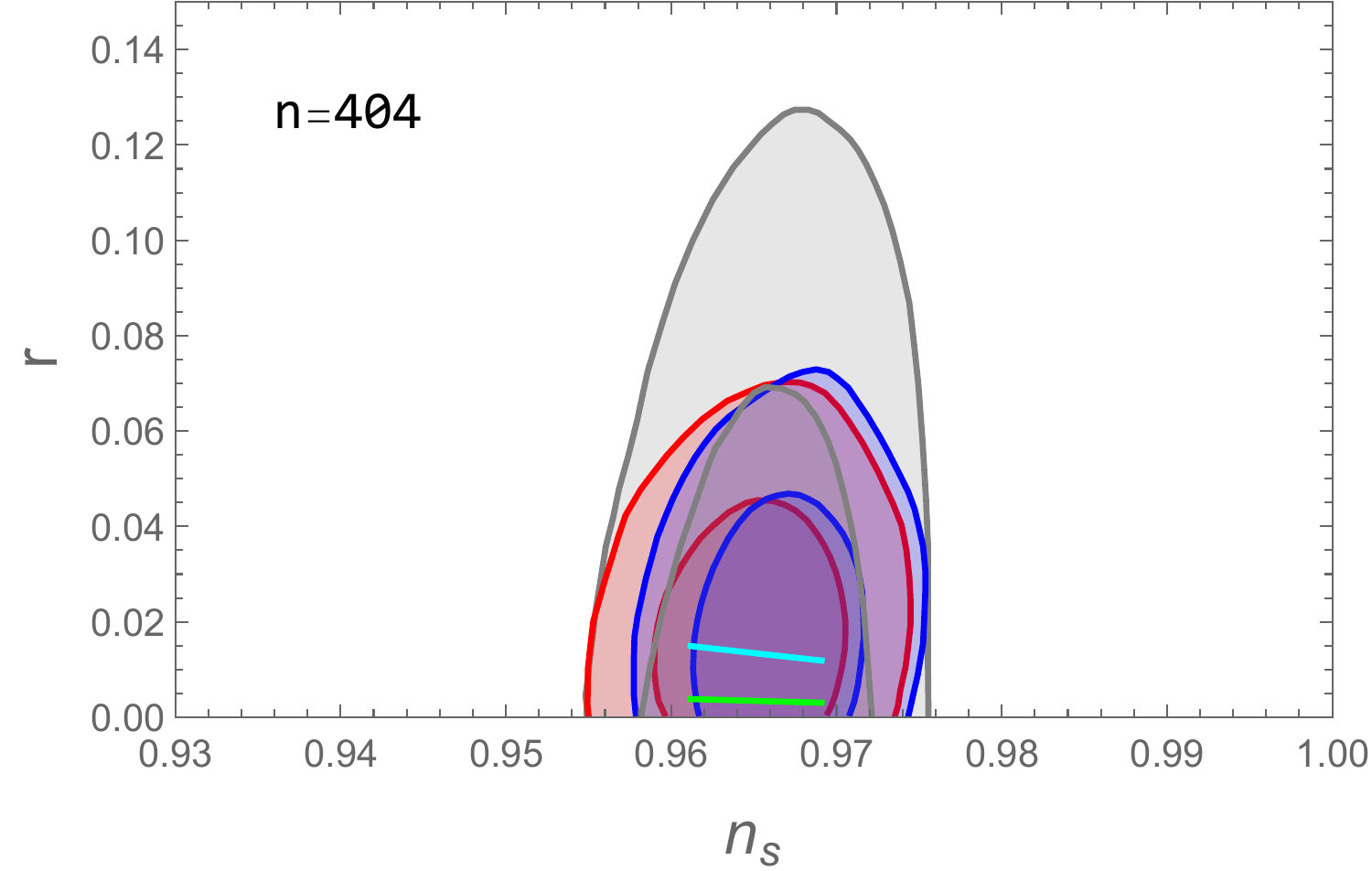}
\includegraphics[width=0.45\textwidth]{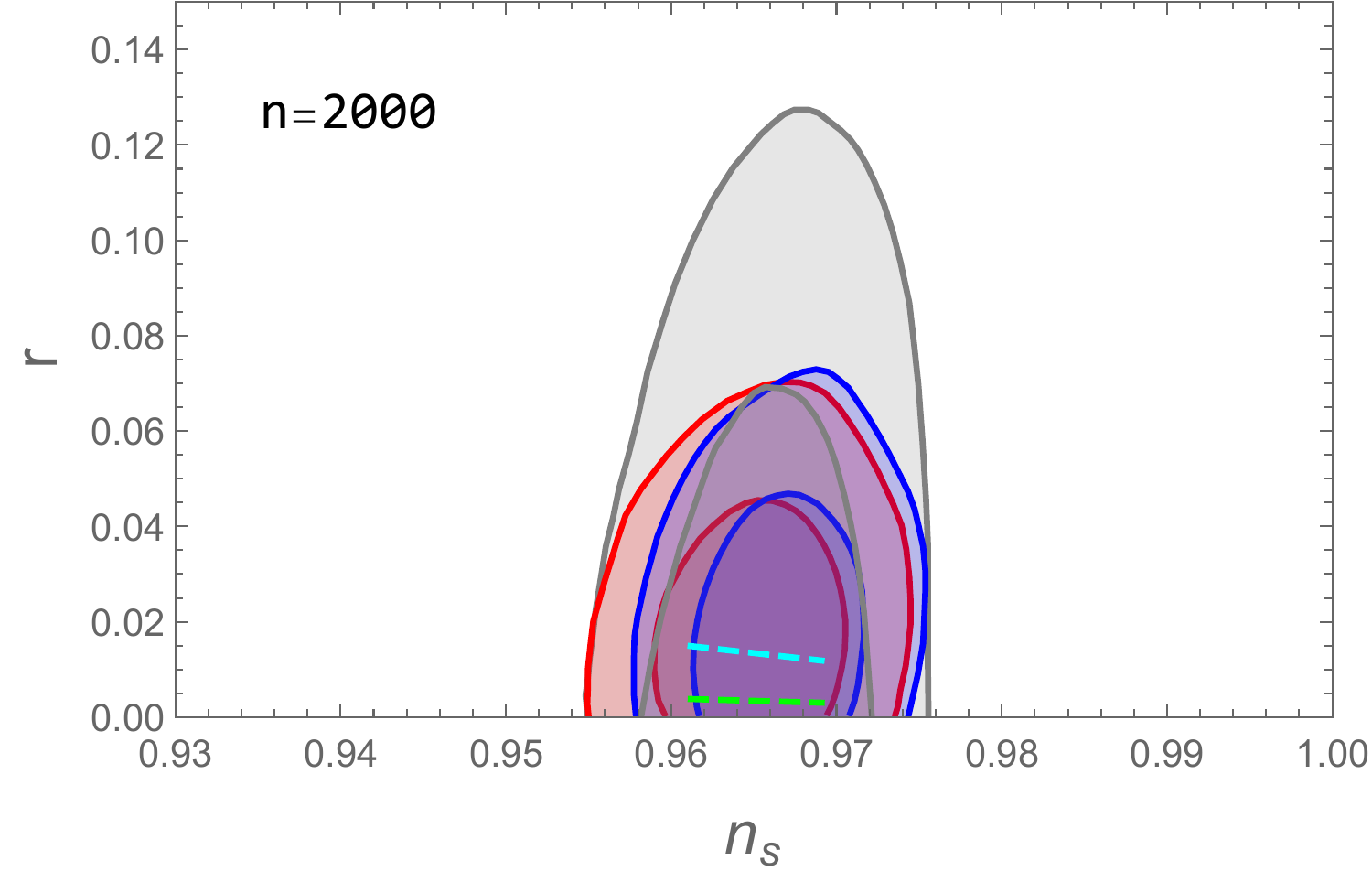}
\end{center}
\caption {ALP inflation predictions in $(n_s,r)$ planes with $n=404$ and $n=2000$.
Gray, red and blue shadows stand for [TT,TE,EE+lowE+lensing], [TT,TE,EE+lowE+lensing+BK15] and [TT,TE,EE+lowE+lensing+BK15+BAO] constraints~ \cite{Akrami:2018odb}.}
\label{Fig:nsr_n}
\end{figure*}

Given the above constraints, we fix $n=404, 2000$ for the rest of analysis. The scalar spectral index $n_s$ and the tensor-to-scalar ratio $r$ are shown in Fig.~\ref{Fig:nsr_n}. The green and cyan lines (dashed) represent that $\phi_*$ are $602.9 \frac{\pi f_\phi}{404}$ and $602.8 \frac{\pi f_\phi}{404}$ ($2998.9\frac{\pi f_\phi}{2000}$ and $2998.8\frac{\pi f_\phi}{2000}$), respectively.
Both $r$ and $n_s$ are almost independent of $n$, and the larger $\phi_*$, the smaller $r$.
To see the specific values more clearly, the upper and lower limits of the corresponding other parameters are listed in Table.~\ref{Tab:parameters}.

\begin{table*}[htb]
\caption{Two sets of benchmark points with $n=404, 2000$ for $f_\phi~ (\rm{GeV})$ and $10^{-15}\Lambda~(\rm{GeV})$, in which the scalar spectral index $n_s = 0.9649 \pm 0.0042$ and $r$ are also presented.}
\begin{center}\label{Tab:parameters}
\renewcommand{\arraystretch}{1.4}
\begin{tabular}{ |c|  c|  c| c|  c|}
\hline
~~~~~~~~~   & $n=404$                           & $n=404$                          & $n=2000$                             & $n=2000$\\
\hline
$\phi_*$   & $603.9 \frac{\pi f_\phi}{404}$     & $603.8 \frac{\pi f_\phi}{404}$   & $2998.9\frac{\pi f_\phi}{2000}$      & $2998.8\frac{\pi f_\phi}{2000}$ \\
\hline
$f_\phi $  & $[10^{21.5388},10^{21.5910}]$      & $[10^{21.5373},10^{21.5895}]$    & $[10^{22.2335},10^{22.5889}]$        & $[10^{22.2319},10^{22.2841}]$  \\
\hline
$10^{-15}\Lambda $ & $[5.62998,5.30161]$        & $[7.99456,7.52826]$              & $[5.62942,5.30107]$                & $[7.99479,7.52848]$\\
\hline
$r$       & $[0.00385,0.00303]$           & $[0.01508,0.01186]$          & $[0.00385,0.00302]$            & $[0.01509,0.01186]$ \\
\hline
\end{tabular}
\end{center}
\end{table*}

\section{Gravitational waves from inflation}\label{GWsIf}

\subsection{GW from quantum fluctuations}

During the ALP-inflation, the quantum fluctuations of inflaton will lead to the perturbations in energy and momentum,
which will transmit to the metric in terms of Einstein's equation,
in which the tensor perturbations of the metric will produce the so-called QF GW.
Generally, the initial tensor power spectrum (spectrum at $\tau = 0$) $P_h(k)$ can be parameterized as a power-law form~\cite{Peiris:2003ff}, i.e.,
\begin{equation}
P_h(k)=rk^2A_R(k_0) \bigg(\frac{k}{k_0}\bigg)^{n_t},
\end{equation}
where $n_t$ is the power index of tensorial spectrum and the value of scalar power spectrum $A_R(k_0)=2.371\times 10^{-9}$ with $k_0=0.002~ \rm{Mpc}^{-1}$~\cite{Ade:2015xua,Ade:2015lrj}.

Furthermore, the present spectrum of energy density in terms of $P_h(k)$ can be written as~\cite{Liu:2015psa}:
\begin{equation}\label{Eq:QF_GWs}
\Omega_{\rm{GW},\rm {QF}}(k)=\frac{P_h(k)}{12H^2(\tau_0)}\times T^2(k,\tau_0)
\end{equation}
where the current Hubble parameter $H(\tau_0)$ is taken as $67.8 \rm{km~s^{-1}~Mpc^{-1}}$, the current conformal time $\tau_0 = 1.41 \times 10^4 ~\rm{Mpc}$,
and the last part $ T(k,\tau_0)$ is the transfer function as follows\cite{Turner:1993vb,Zhang:2005nw,Giovannini:2009kg,Chongchitnan:2006pe,Kuroyanagi:2009br}:
\begin{equation}
T(k,\tau_0)=\frac{3\Omega_m j_1(k\tau_0)}{k\tau_0}\sqrt{1.0+1.36\frac{k}{k_{eq}}+2.50\bigg(\frac{k}{k_{eq}}\bigg)^2}
\end{equation}
where matter density $\Omega_m=0.308$, the spherical Bessel function of the first kind $j_1(k\tau_0)\sim\frac{1}{\sqrt{2}k\tau_0}$, and $k_{eq} = 0.073 \Omega_m h^2 ~\rm{Mpc}^{-1}$ is the wavenumber corresponding to the mode that entered the horizon at the equality of matter and radiation.

\begin{figure*}[t]
\begin{center}
\includegraphics[width=0.45\textwidth]{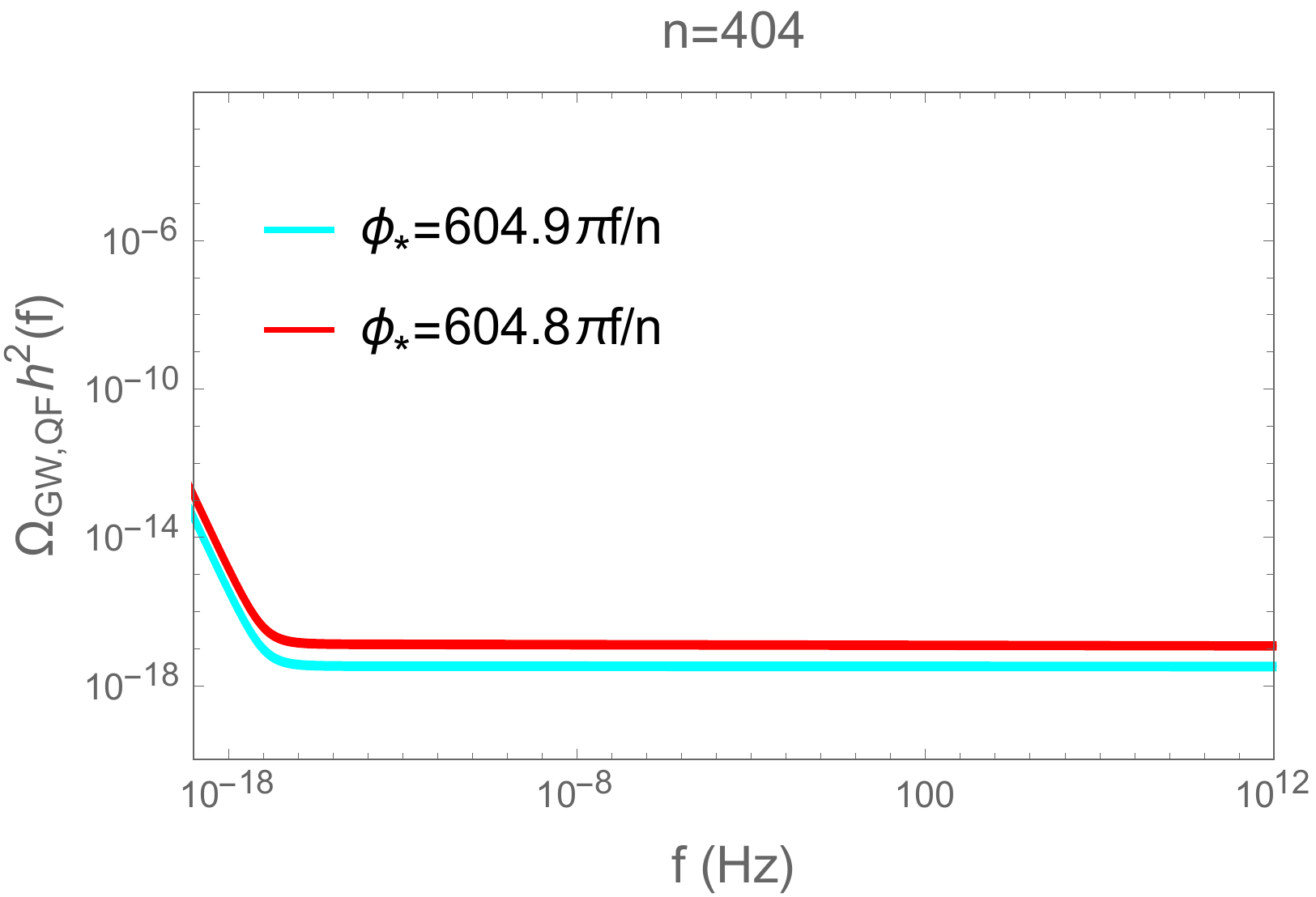}
\includegraphics[width=0.45\textwidth]{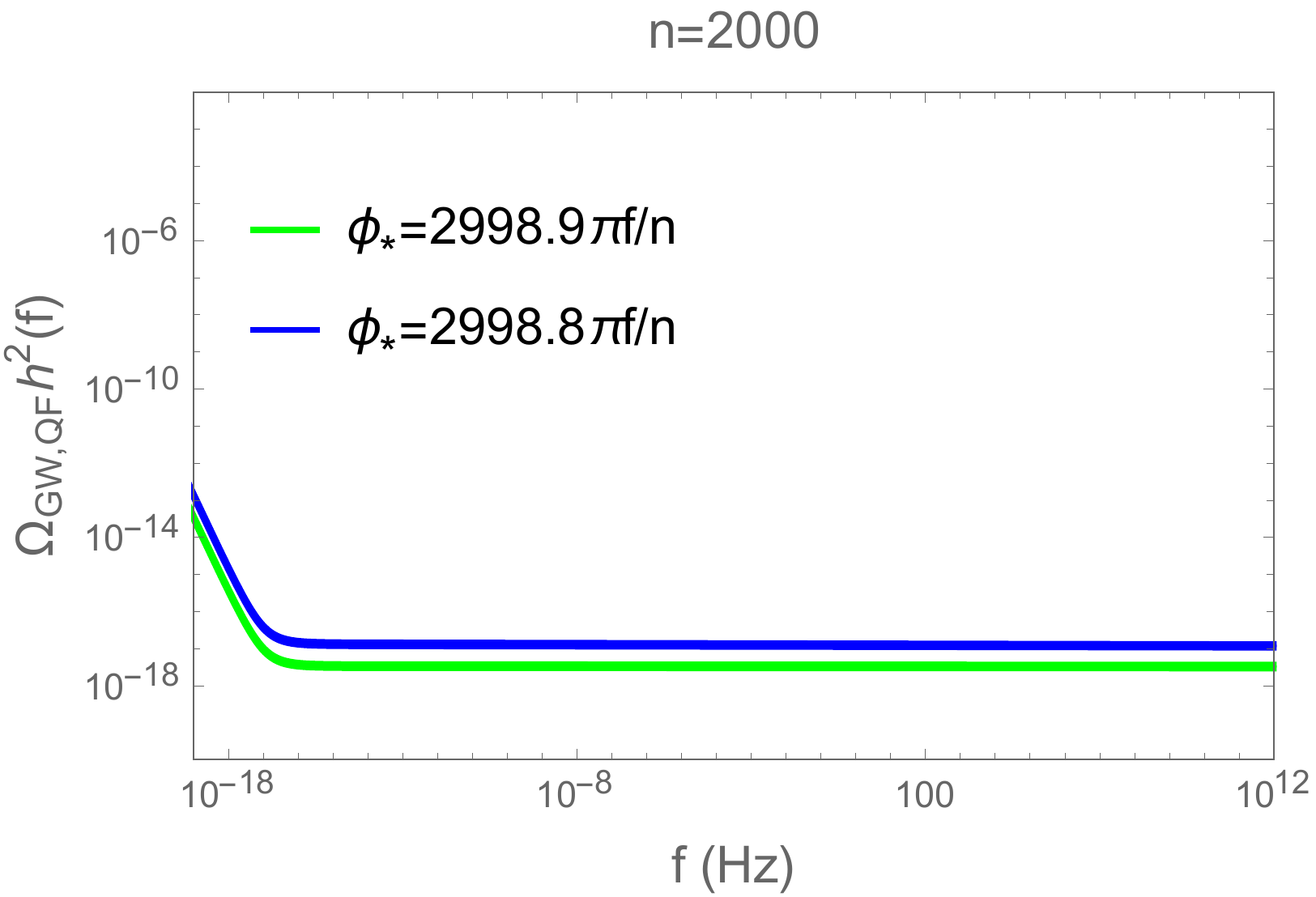}
\end{center}
\caption{The spectrum of GW from the quantum fluctuations during the ALP inflation.}
\label{Fig:GW_QF_nall}
\end{figure*}

Applying the input parameters in Table.~\ref{Tab:parameters} and Formula~\ref{Eq:QF_GWs}, we show the GW from the QF during the ALP inflation in Fig.~\ref{Fig:GW_QF_nall}, in which the left and right parts stand for the QF GW with $n=404$ and $n=2000$, respectively. The QF GW first decreases with the increase of frequency $f$, and then remains almost constant from frequency $f\sim10^{-16}~ \rm{Hz}$. Those two lines with $\phi_*=604.9\pi f/n$ and $\phi_*=2998.9\pi f/n$ ($\phi_*=604.8\pi f/n$ and $\phi_*=2998.8\pi f/n$) almost overlap each other due to the same value of tensor-to-scalar ratio $r$ that are shown in Table.~\ref{Tab:parameters}, which also implies the QF GW are not sensitive to the parameter $n$. When we fix the parameter $n$, the QF GW will increase with the of decrease $\phi_*$.

\subsection{GW from phase transitions}

As we emphasized earlier, in our model, ALP inflation exited through bubble nucleations rather than the breach of slow-roll conditions.
Numerical calculation of two bubble collisions and more bubble collisions can be referred to Ref.~\cite{Kosowsky:1992vn} and Ref.~\cite{Huber:2007vva}, respectively. While the bubble nucleations will lead to PTs that further produce PT GW. The GW for a single PT can be expressed as~\cite{Kosowsky:1992vn}:
\begin{equation}\label{mth profile}
\Omega_m h^2(f)=\left\{
\begin{array}{ll}
\Omega h^2(f_m){\left(\frac{f}{f_{m}}\right)}^{2.8}\quad ~~ (f\leq  f_{m}), &\\
\Omega h^2(f_m){\left(\frac{f}{f_{m}}\right)}^{-1.8}\quad  (f\geq  f_{m}), &
\end{array}
\right.
\end{equation}
where the peak frequency today $f_m= \frac{3\times 10^{-8}}{\chi}e^{-((n-2)/2-m-1)\chi}{\left(\frac{g_{\ast}}{100}\right)}^{1/6}\left(\frac{T_e}{1~{\rm GeV}}\right)$
with the temperature of the universe at the end of ALP inflation being $T_e=\frac{1}{1-e^{-\chi}}{\left(\frac{30\delta \epsilon}{g_{\ast}\pi^2} \right)}^{1/4}$.
The total number of relativistic degrees of freedom is taken to be $g_{\ast} \simeq 100$. And the amplitude $\Omega h^2(f_m)$ at this peak frequency have the following form:
\begin{eqnarray}
\Omega h^2(f_m)=10^{-6}(1-e^{-\chi})^4 \chi^2
e^{-4((n-2)/2-m-1)\chi}{\left(\frac{100}{g_{\ast}}\right)}^{1/3} .
\end{eqnarray}

Thus, the total phase transition energy density of the gravitational waves is given by a sum of the multiple tunneling events,

\begin{equation}\label{Eq:PT_GWs}
\Omega_{\rm{GW},\rm {PT}} h^2(f) = \sum_{m=1}^{(n-2)/2-1} \Omega_m h^2(f).
\end{equation}

\begin{figure*}[ht!]
\begin{center}
\includegraphics[width=0.45\textwidth]{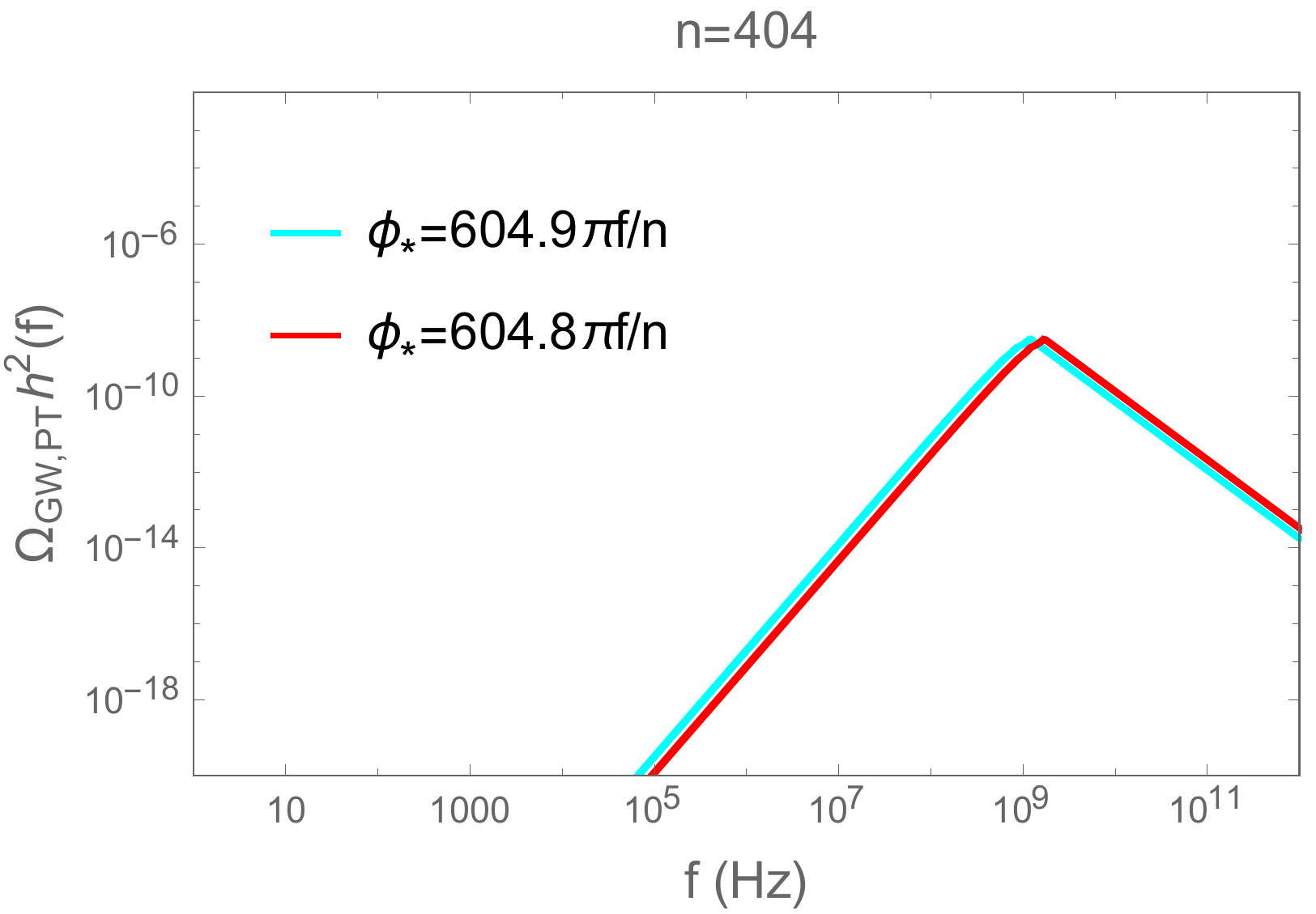}
\includegraphics[width=0.45\textwidth]{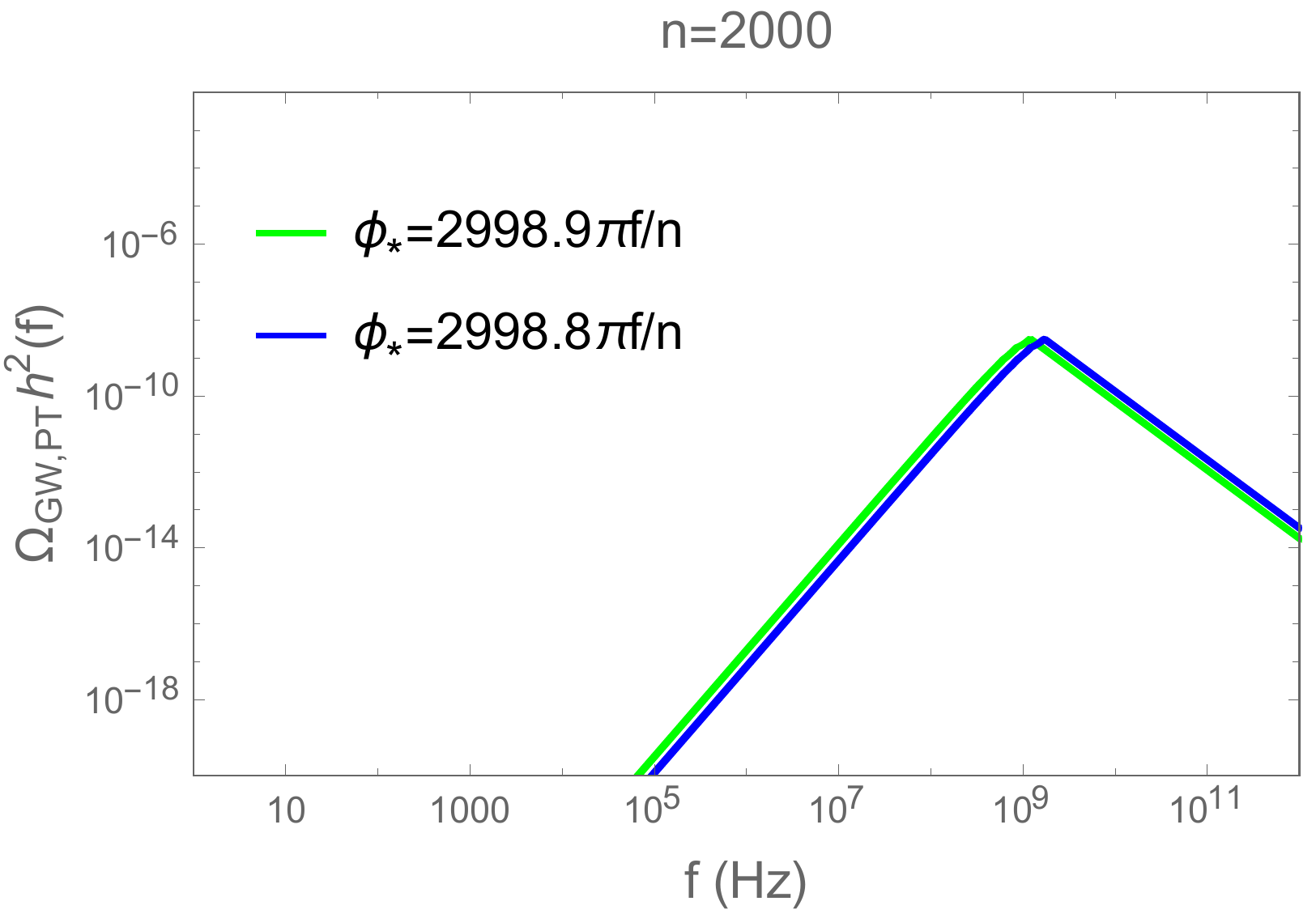}
\end{center}
\caption{The spectrum of PT GW from the ALP inflation.}
\label{Fig:GW_PT_nall}
\end{figure*}

We show the PT GW in Fig.~\ref{Fig:GW_PT_nall}, in which the left and right parts stand for the PT GW with $n=404$ and $n=2000$, respectively.
The PT GW have almost identical peaks and shapes for different $n$, which also implies the PT GW is insensitive to $n$, but the energy spectrum $\Omega_{GW,PT}h^2$ sharp is shifted about five times to the left as $\phi_*$ increases.
We can see that the GW produced by PT during inflation is high-frequency GW. Compared to the previous QF GW, it dominates the low-frequency regions, while PT produces GW that will dominate the high-frequency regions.

\subsection{Total GW}

After acquiring the PT GW and QF GW during the ALP inflation, we can now analyze the total GW.
According to the Eqs.~(\ref{Eq:QF_GWs}) and (\ref{Eq:PT_GWs}), it can be expressed as follows:
\begin{equation}\label{total-spectrum}
 \Omega_{\rm{GW},\rm {Tot}}h^2=\Omega_{\rm{GW},\rm {PT}} h^2+\Omega_{\rm{GW},\rm {QF}}h^2.
\end{equation}

\begin{figure*}[ht!]
\begin{center}
\includegraphics[width=0.45\textwidth]{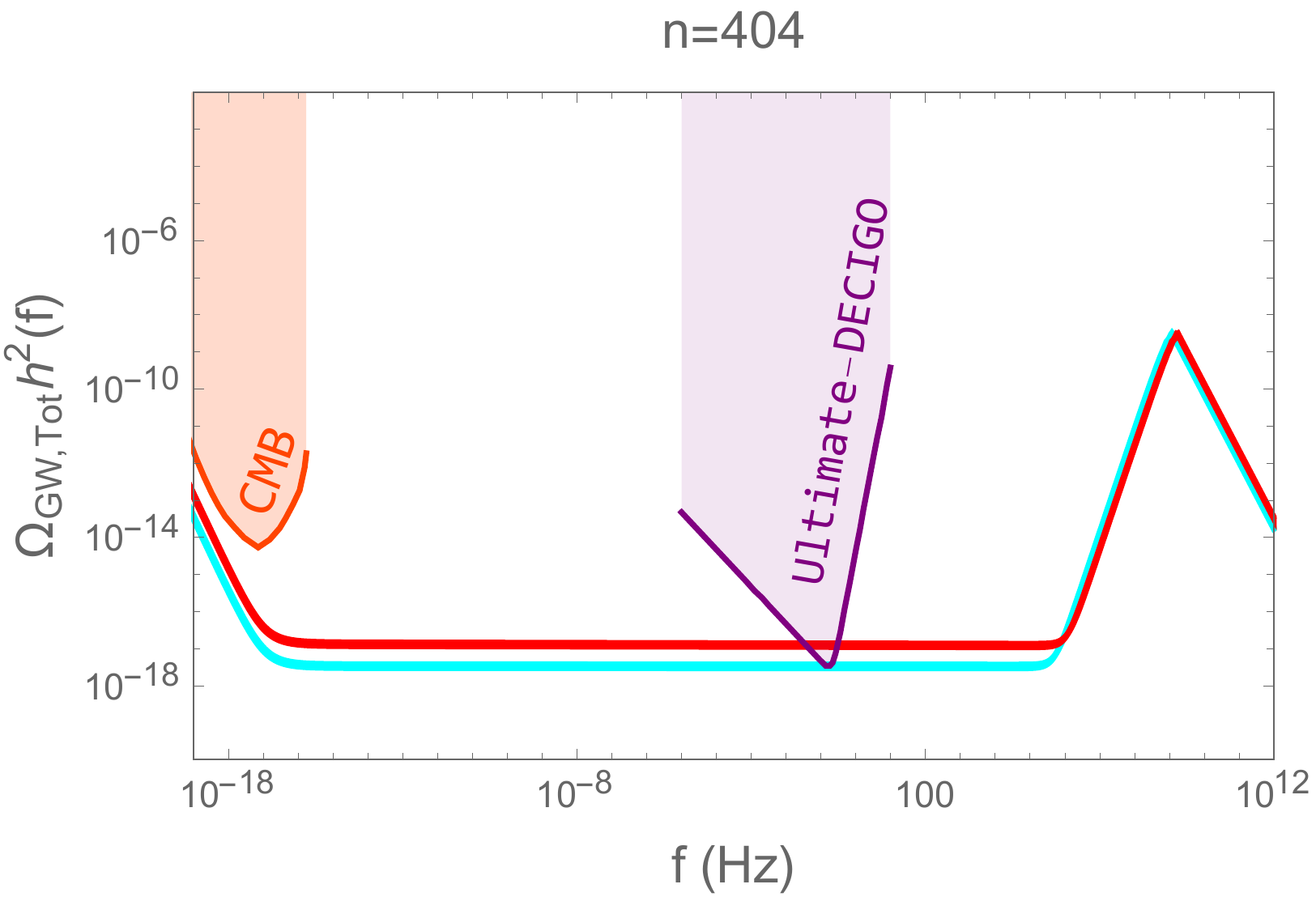}
\includegraphics[width=0.45\textwidth]{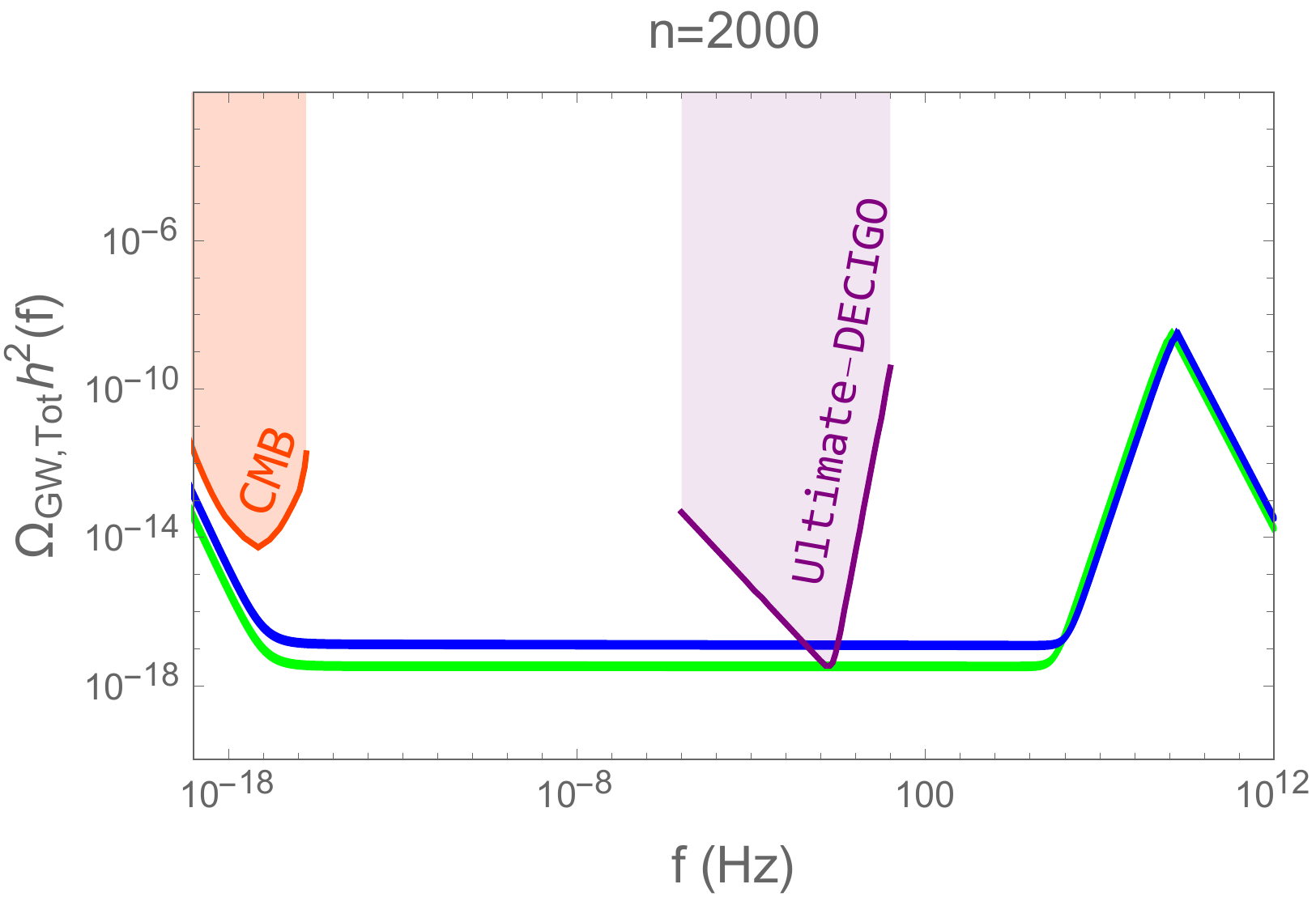}
\end{center}
\caption{The spectrum of total GW from the ALP inflation. The CMB observation make a limit in the low-frequency~\cite{Lasky:2015lej,Sepehri:2016jsu}, while the detector Decigo~\cite{Yagi:2011wg} can provide a test for the GW in the high-frequency.}
\label{Fig:GW_Tot_nall}
\end{figure*}

The total GW are shown in Fig.~\ref{Fig:GW_Tot_nall}, we find that at ultra-low frequency $f\sim(10^{-19},10^{-17})~\rm{Hz}$, all the predictions of the GW meet the CMB limits~\cite{Lasky:2015lej,Sepehri:2016jsu}, while near the frequency $f\sim(10^{-1})~\rm{Hz}$, the detector Decigo~\cite{Yagi:2011wg} can search for all of the GW scenarios that we predict.
For the extremely high-frequency $f\sim(10^{9},10^{12})~\rm{Hz}$, the total GW $\Omega_{\rm {Tot}}h^2$ of our prediction has been reached $\sim10^{-10}~\rm{Hz}$, which is expected to be searched on the $3\rm{DSR}$ detector proposed by Chongqing University~\cite{Li:2003tv,Tong:2008rz}.

\section{Summary}\label{Summary}

In this paper, we investigate the GW generated by the MNI that exits from the PT. This change makes this model better match the CMB observation data than the usual slow-roll exit scenario. The GW will consist of two parts, i.e., GW is generated by QF during inflation and by PT finished the inflation, and it will bring a sizable GW that may be detected by the future GW detectors.

As the PT events need more than $200$, therefore, the model parameter $n$ should not less than $404$. We investigate two scenarios ($n=404, 2000$), and find that generated GW is very insensitive to $n$, which can be obtained from all the GW figures. The GW generated by QF dominates the low-frequency part, while the GW generated by PT dominates the high-frequency part. At $f\sim10^{-1}\rm{Hz}$, the generated GW can be detected by Decigo~\cite{Yagi:2011wg}, while in the extremely high-frequency part, it is expected to be detected by $3\rm{DSR}$ designed by Chongqing University.

\hspace{2cm}

\noindent{\bf Acknowledgement}:

This work has been supported by the National Science Foundation of China under Grant No.12047564, the Fundamental Research Funds for the Central Universities under Grant No. 2020CDJQY-Z003, and the China Postdoctoral Science Foundation under Grant No.~2019TQ0329 and No.~2020M670476.

\bibliographystyle{arxivref}

\end{document}